\newcommand {\be}{\begin{equation}}
\newcommand {\ee}{\end{equation}}

\newcommand {\bea}{\begin{eqnarray}}
\newcommand {\eea}{\end{eqnarray}}

\documentstyle[aps,prl,epsf]{revtex}
\draft

\begin{document}

\twocolumn[\hsize\textwidth\columnwidth\hsize\csname@twocolumnfalse%
\endcsname

\title{
A new paradigm for two-dimensional spin-liquids}
\author{R. R. P. Singh$^1$, O. A. Starykh$^2$ and P. J. Freitas$^1$}
\address{$^1$Department of Physics, University of California, Davis, 
California 95616\\
$^2$ Department of Physics, University of Florida, Gainsville, FL 32611}
\date{\today}
\maketitle{}

\begin{abstract}
Motivated by the geometry of the materials 
Na$_2$Ti$_2$As$_2$O and Na$_2$Ti$_2$Sb$_2$O,
we study a square-lattice Heisenberg antiferromagnet, with spins
located at the bond-centers. The largest
exchange constant $J$ couples neighboring spins 
in a given row or column. 
This leads to a mesh of isolated spin-chains running along
the X and Y axes.
A weaker exchange constant $J^\prime$
couples the nearest-neighbor spins on the lattice. 
Classically, $J^\prime$ fails to fix the relative
spin orientation for different chains and hence
the ground state is highly degenerate.
Quantum order by disorder effect is studied by spin-wave
theory and numerical methods. It is shown that a 4-sublattice
order is favored by quantum fluctuations. However, several
arguments are presented that suggest that the ground state of
the system remains disordered, thus providing us with a paradigm for
a two-dimensional spin-liquid.
\end{abstract}
]

In recent years many new materials have been discovered which
exhibit novel magnetic behavior. Various aspects of quantum magnetism,
including quantum critical phenomena and existence of spin-disordered
ground states with spin-gaps 
have been observed in 
a variety of Cuprates, Germanates, Vanadates and other low-dimensional
materials.
One interesting fact that has come to light in these studies
is that the geometrical arrangement of
the transition metal and Oxygen ions can have a dramatic impact
on the underlying microscopic spin-Hamiltonian and hence on the
macroscopic magnetic properties of the system. For example in various
Cuprates,
the Cu-O-Cu bond angle is crucial for determining
the effective exchange constant between Copper spins.
Thus, Strontium Cuprates with certain stoichiometry behave
as virtually decoupled spin-ladders \cite{srcuo},
even though the separation of the Copper spins between neighboring
ladders maybe smaller than their separation within a given ladder.
In the CuGeO$_3$ \cite{cugeo} and CaV$_n$O$_{2n+1}$ \cite{cavo} 
it is also believed
that superexchange between the transition metal ions is
mediated by Oxygen and because of the geometry of various bond angles
and occupied orbitals the second neighbor interactions are substantial
compared to nearest neighbor ones. This leads to various
interesting quantum phase transitions and spin-gap behavior 
in these materials.

Here, we consider a Heisenberg model with spins
at the bond-centers of a square lattice:
\be
{\cal H}=J\sum_{<i,j>} \vec S_i\cdot\vec S_j
+J^\prime\sum_{<i,j>} \vec S_i\cdot\vec S_j,
\label{Ham}
\ee
with $J^\prime<<J$. The interactions are shown in Fig. 1.
The exchange $J$ couples neighboring spins in a given row or column,
whereas $J^\prime$ is the nearest neighbor coupling between rows and
columns. In
the absence of $J^\prime$ the system consists of a square mesh
of decoupled spin-chains running along the X and Y axes. 
The motivation for studying such a model comes from the
materials Na$_2$Ti$_2$Sb$_2$O ( and also Na$_2$Ti$_2$As$_2$O)
\cite{adam,kauz}.
These layered Titanates consist of planes of (Ti$_2$Sb$_2$O)$^{2-}$,
where the Oxygen atoms form a square-lattice and the Titanium atoms
sit at the bond centers of the lattice. The Antimony atoms
sit above and below the centers of the elementary squares. The Titanium
atoms carry spin-half. 
It is evident from the geometry that if the dominant exchange
interaction is provided by a direct overlap between the Titanium
orbitals, this material would behave as a nearest-neighbor Heisenberg
model. If on the other
hand the dominant interaction is mediated by Oxygen orbitals
it would result in the largest interaction between neighboring spins 
in a given row or column, thus leading to our Eq.~\ref{Ham}
\cite{Goodenough}.
Finally, if the dominant exchange 
is mediated by the p-orbitals of Antimony, 
the Hamiltonian in Eq.~\ref{Ham} could still be  appropriate, the
lattice
being dual to the one shown.
The magnetic properties of this material are currently under
investigation
\cite{kauz}.
The rest of the paper will be devoted to a study of this model
Hamiltonian.

\begin{figure}[htb]
\setlength{\unitlength}{1.0in}
\begin{picture}(2.0,2.5)(0.1,0)
\put(0.45,0.3){\epsfxsize=2.5in\epsfysize=2.0in\epsfbox{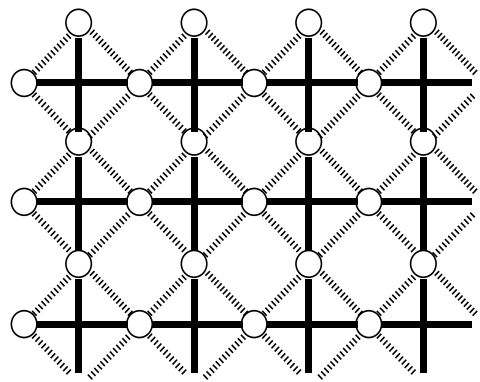}}
\end{picture}
\caption
{The exchange interactions $J$ and $J^\prime$ in our model,
shown by solid lines and broken lines respectively.}
\end{figure}

We begin by studying this model in the linear spin-wave
approximation. As finding the correct local spin configuration
is based on energetic considerations, we expect the spin-wave theory to
be at least qualitatively correct. Notice that the problem at hand
has `double' degeneracy in the classical limit: it is evidently
degenerate
with respect to relative angle $\theta$ between quantization axes on
vertical and horizontal chains, as well as with respect to relative
orientation $\phi$ of quantization axes on the neighboring $parallel$
chains. Based on the fact that Hamiltonian (\ref{Ham}) classically has
no
spiral ordering, and on insight from a somewhat analogous problem of
coupled planes \cite{yhs}, we restrict variation in $\phi$ to two
possible values: $\phi=0$ (ferromagnetic ordering of neighboring
parallel
chains) and $\phi=\pi$ (antiferromagnetic one). This degree of freedom
is then naturally represented by Ising-type discrete variable
$\tau_{i,i+1}=1 (\phi=0)$ or $\tau_{i,i+1}=0 (\phi=\pi)$, defined
for each pair of chains $(i,i+1)$. 
The calculations are simplified
greatly by choosing quantization axes on all sites such that the
ordering is ferromagnetic.
This is achieved by a unitary  transformation 
parametrized by an angle $\theta$ and a set of $(\tau_{i,i+1})$.

We now integrate out spins on the horizontal chains to find 
an effective Hamiltonian for the remaining spins. This is
achieved by representing spins on vertical (horizontal) chains in
terms of Holstein-Primakoff bosons $a (b)$, and doing perturbation
expansion in $J^{\prime}/J$. First nonzero contribution is of the type
$J^{\prime~2} \sum_{e,e^{\prime}} S^{\alpha}_{i + e} (a) S^{\beta}_
{j+e^{\prime}} (a) < S^{\alpha}_i (b)  S^{\beta}_j (b)>_b$,
where averaging is with respect to the bare Hamiltonian of horizontal
chains
and $S_{i + e}$ spins belong to vertical chains,
i.e. $J^{\prime}$ acts along the $<i,i+e>$ links. The average is nonzero
when $i$ and $j$ belong to the same horizontal chain and $\alpha=\beta$.
It is seen that effective interaction between spins on vertical chains 
is time-dependent,
but strongest contribution comes from static ($\omega_n=0$) part of the
average.
Denoting $J_{\perp}^{\alpha}(i-j)=J^{\prime~2} < S^{\alpha}_i (b)
 S^{\beta}_j (b)>_b$ and
$\hat{J}_{\perp}^{\alpha}(r)=J_{\perp}^{\alpha}(r-1)
+ J_{\perp}^{\alpha}(r)$ we find that geometry dictates the
following coupling between
remaining parallel chains $n$ and $m$,
\be
H_{nm}^{\perp}= \sum_{i \alpha} \hat{J}_{\perp}^{\alpha}(n-m)
S^{\alpha}_{n,i}\left( 2S^{\alpha}_{m,i} +
S^{\alpha}_{m,i+1} + S^{\alpha}_{m,i-1}\right).
\label{eff-ham}
\ee
Notice that long-range-ordered part of $<S^{\alpha}_i (b) S^{\beta}_j
(b)>_b$,
if any, cancels out due to AFM correlations along the chains,
and the effective interaction decays at least as $(n-m)^{-2}$.
It is also important to realize that $\hat{J}_{\perp}^{\alpha}$ is
anisotropic
in spin space, in particular $\hat{J}_{\perp}^{x(z)} \sim
\cos^2{\theta}$.
Notice that classically the system remains degenerate because 
$< 2S^{\alpha}_{m,i} +S^{\alpha}_{m,i+1} + S^{\alpha}_{m,i-1}>=0$.
Given the fast decay of the induced inter-chain interaction with the
distance
we restrict ourselves to the strongest interaction, 
$\hat{J}_{\perp}^{\alpha}(1)$, between neareast chains. To find
corrections
to ground state energy due to this coupling we next 
treat $\hat{J}_{\perp}$ as a perturbation
(see Ref.\cite{yhs} for description of the procedure). 
After long algebra one finds quantum correction to 
the ground state energy (per chain) of $M$ parallel chains of length
$N$:
\be
\delta E_{G.S.}= 
- \frac{1}{M}\sum_{q_y} \epsilon(q_y) C (J^{\prime}/J)^4 \cos^2{\theta}
\sum_{i=1}^{M} (1 - 2\tau_{i,i+1}),
\label{structure}
\ee
where 
$\epsilon(q)=2JS|\sin{q}|$ is the single AFM chain 
dispersion in spin-wave theory, 
$C=(<S^x_0 S^x_0> + <S^x_0 S^x_1>)^2/64$, and explicit form of 
$\hat{J}_{\perp}$was used. 
Note that correction is down by $1/S$ factor, showing its quantum
origin. Thus the energy is
minimized by choosing all $\tau_{i,i+1} =0$ (AFM configuration) and 
$\cos^2{\theta}=1$. This result agrees with well-known tendency of
quantum
fluctuations to favor colinear structures \cite{j1j2,chandra}.

Having found the ground state configuration of spins, we can
determine the spin-wave dispersion
and calculate the reduction in sublattice magnetization due to
quantum fluctuations. This reduction turns out to be divergent due
to a remaining artificial degeneracy of the spin-wave spectra.
Similar divergency  
arises in a quantum order by disorder calculations on other systems and 
is known to be removed by the higher-order quantum corrections to the
spin-wave spectra \cite{chubukov}.
This technically difficult calculation has not been done.
We would like
to stress that the ground state structure found in (\ref{structure})
is determined by short-range spin correlations (i.e. by the correlations
within the correlation length range) for which spin-wave 
approximation should work even if the sublattice magnetization vanishes.

Let us now take a closer look at the elementary excitations of the
single
chain. It is convenient to perform ``staggering'' of the spin
configuration
so that N\'eel ordering along the chain corresponds to the ferromagnetic one
in the new representation. In this representation an elementary 
excitation of the chain is a domain
wall (spinon) separating ferromagnetic domains of different orientation.
The energy of a single domain wall is $J_z$.
Consider now unfrustrated coupling $J_{\perp}$ between nearest
spins on neighboring parallel chains:  however small $J_{\perp}$ is, it
immediately leads to the suppression of spinons because the
energy costs is proportional to $J_{\perp}\times$(length of the domain)
and diverges in the thermodynamic limit. This is an intuitive reason
for the stabilization of LRO in the system of unfrustrated coupled
chains \cite{schulz}. But this is not true for our Hamiltonian
(\ref{eff-ham}),
where each spin is coupled to the zero-spin combination of spins
on the neighboring chains. Thus domain wall excitations of the single
chain
seems to survive in the presence of nonzero $J^{\prime}$, 
hinting to the possibility of the absence of LRO.

The effect of the ``mixing'' interchain coupling $J^{\prime}$ can also
be taken 
into account
in the disordered phase, where the interspin interaction is isotropic. 
We write the partition function of the
Hamiltonian (1) in the interaction representation, where independent
chains are considered as unperturbed system and interchain coupling
$J^{\prime}$ as a perturbation \cite{tremblay}. Performing trace
over the horizontal chains first
one finds effective interaction between spins on the vertical chains
proportional to $(J^{\prime}({\bf k}))^2 G^{(0)}(k_x,\omega_n)
S^{\alpha}
({\bf k}, \omega_n) S^{\alpha}({\bf k}, \omega_n)$, similar to our
previous
spin-wave calculations. Here $G^{(0)}(k_x,\omega_n)$ is the 
rotationally-invariant one-dimensional
spin Green's function in Matsubara representation,
and $J^{\prime}({\bf k})=J^{\prime}\cos{k_x \over 2} \cos{k_y \over 2}$. 
Decoupling spins on different chains via a
Hubbard-Stratonovich transformation \cite{tremblay},
the 2D Green's function of the spins on vertical
chains becomes,
\be
G(k_x,k_y,\omega_n)=\frac{G^{(0)}(k_y,\omega_n)}{1 - (J^{\prime}({\bf
k}))^2
 G^{(0)}(k_x,\omega_n) G^{(0)}(k_y,\omega_n)}.
\label{green}
\ee
The known form of $G^{(0)}(k_x,\omega_n)$ \cite{subir} implies
that
($i$) the singularity at ${\bf k}_0=(\pi,\pi)$ remains unchanged 
because $J^{\prime}({\bf k}_0)=0$,
($ii$) the uniform susceptibility is 
$\chi=\chi_0/[1 - (\chi_0 J^{\prime})^2]$,
where $\chi_0=1/(\pi^2 J)$ is susceptibility of single chain,
and ($iii$) interchain coupling comes into play below $T^{*} \sim
J^{\prime}$.
Within this approximation, the two-dimensional
system remains in the disordered critical
state at $T=0$, and in the quantum-critical one at finite temperatures.
This result should also be viewed with some caution due to the large
degeneracies in the underlying problem which we do not know yet how
to take into account completely.

We now turn to numerical studies. Because of multiple energy scales,
the development of short-range order at high temperatures is not
necessarily indicative of the order for $T\sim J^\prime <<J$.
Thus we need to study the ordering tendency directly at low
temperatures.
As $J^\prime/J<<1$, the coupling between the
local order parameters on the neighboring parallel
chains can be gotten by perturbation theory.
We study the following correlation function between 
neighboring vertical chains:
\be
c_{12}=<\sum_{i}(-1)^i\vec S_{1,i}\cdot \sum_j (-1)^j\vec S_{2,j}>,
\ee
here $S_{1,i}$ represents spins on first vertical chain and $S_{2,i}$
represents spins on second vertical chain. A value of $c_{12}>0$ would
imply ferromagnetic ordering whereas $c_{12}<0$ would imply 
antiferromagnetic ordering. 
The leading order contribution to $c_{12}$ 
requires the interactions to be mediated by
at least two horizontal chains.
To make the perturbation theory numerically
tractable, we confine ourselves to finite chains along vertical and
horizontal directions ( with periodic boundary conditions).
For four spins in each of the chains, we
can carry out perturbation theory using series expansion methods
\cite{series}.
We find that $c_{12}\approx -9.
\times 10^{-6}(J^\prime/J)^6$, which is negative confirming the tendency
of the parallel chains to align antiferromagnetically.
We have also used Lanczos methods
to calculate this correlation function with upto 6-spins in each chain
for $|J^\prime/J|\le 0.3$.
In all cases the coupling remains antiferromagnetic and very weak.
For the $6$-spin system $c_{12}$ appears to scale 
as $(J^\prime/J)^4$ as expected
from general arguments and spin-wave theory presented earlier.
Complete diagonalization of a 32-site system is currently in progress.

Assuming the local ordering pattern obtained in spin-wave theory, we can
now
investigate the question of long-range order by an Ising expansion
for our original Hamiltonian.
We write the Hamiltonian as
${\cal H}(\lambda)=H_0+\lambda H_1$,
where,
$H_0$ consists of an Ising Hamiltonian, whose two ground states are
ordered in the four-sublattice pattern and $H_1={\cal H}-H_0$, with
${\cal H}$ being the full Hamiltonian in Eq.~\ref{Ham}. We
develop expansions for the sublattice magnetization in terms of
the expansion parameter $\lambda$. If the system has
long-range order, the expansion should converge upto $\lambda=1$,
whereas if it is disordered there should be a critical point
at $\lambda<1$. Without explicit calculations,
we can make the following observation: because of the cancellation of 
the effective field from one chain to another, the purely 
one-dimensional graphs remain unaffected by the coupling between
the chains. We know that these
add up to zero sublattice
magnetization at the Heisenberg point. The effects of the
other graphs is to further reduce order, thus moving the Heisenberg
system
into a quantum disordered phase.
We have carried out these
expansions to order $\lambda^8$ for a range of $J^\prime/J$ values.
The analysis suggests critical values less than unity,
implying a disordered ground state for the Heisenberg limit.

Additional insight can be gained by studying the 
$J^{\prime} \gg J$ limit of the model, where the N\'eel
state on $45$-degrees rotated lattice is stabilized by the
$J^{\prime}$ interaction. As $J$ increases from zero,
quantum fluctuations grow stronger and at $(J/J^{\prime})_{crit}=0.76$ 
(within linear spin-wave approximation)
finally destroy long range order, analogous to 
$J_1 - J_2$ model \cite{j1j2}. That these two systems are very
similar is also supported by the fact that the critical value
of frustrating interaction $J$ quoted above is exactly twice
the corresponding critical value of $J_1 - J_2$ model. Given
this analogy, it is tempting to speculate that disordered phase at
$J/J^{\prime} \ge 0.76$ is a spontaneously dimerized
one\cite{dimer,series},
with vertical and horizontal chains formed by the exchange $J$
being in the valence-bond-type state. However, extensions of these
results to $J/J^{\prime}>1$ is problematic as at $J=J^\prime$ this
system classically has a finite ground state entropy \cite{Henley}
and may change character at that point.
Clearly more work is needed to understand the nature of the 
disordered phase.
It is very interesting to ask what will happen upon introducing of a
mobile charge carriers into the chains. We would like to point out
apparent similarities of this problem with a very recent study of a square
mesh of conducting horizontal and vertical stripes \cite{castro}.

In conclusion, we have identified a new frustrated layered Heisenberg
model, which could have a two-dimensional
spin-liquid ground state. The magnetic properties of the materials 
Na$_2$Ti$_2$As$_2$O and Na$_2$Ti$_2$Sb$_2$O
maybe represented by such a model.

We would like to thank S. Kauzlarich and A. Axtell
for bringing these materials to our attention, and
I. Affleck, A. Chubukov, 
S. Sachdev and M. Zhitomirsky for discussions and suggestions.
This work is supported in part by the National Science Foundation Grant
DMR-9616574, the Campus Laboratory Collaboration of
the University of California, and NHMFL.

\end{document}